# A nanostructured surface increases friction exponentially at the solid-gas interface

Arindam Phani[1*], Vakhtang Putkaradze[1,2*], John E. Hawk[1], Kovur Prashanthi[1], Thomas Thundat[1*]

[1]*Department of Chemical and Materials Engineering,* [2]*Department of Mathematical Statistical Sciences, University of Alberta, Edmonton, Alberta T6G 1H9, Canada*

*Correspondence to: phani@ualberta.ca, putkarad@ualberta.ca, thundat@ualberta.ca*

According to Stokes' law, a moving solid surface experiences viscous drag that is linearly related to its velocity and the viscosity of the medium. The viscous interactions result in dissipation that is known to scale as the square root of the kinematic viscosity times the density of the gas. We observed that when an oscillating surface is modified with nanostructures the experimentally measured dissipation shows an exponential dependence on kinematic viscosity. The surface nanostructures alter solid-gas interplay greatly, amplifying the dissipation response exponentially for even minute variations in viscosity. Nanostructured resonator thus allows discrimination of otherwise narrow range of gaseous viscosity making dissipation an ideal parameter for analysis of a gaseous media. We attribute the observed exponential enhancement to the stochastic nature of interactions of many coupled nanostructures with the gas media.

**Introduction.** Solid-gas interface interactions are central to physical processes such as adsorption, catalysis, oscillatory dynamics, stochastic interactions relevant to thermodynamics, atomic and molecular manipulation, viscous drag on nanoparticles and many other fields[1–6]. Dissipative effects from viscous drag, originating from the solid-gas interactions, play an important role in recent advances in micro-nano resonator technology[7–14]. Viscosity in general is related to momentum transfer originating from collisions of fluid molecules with surfaces. Thus, viscous friction[15], is always present resulting in energy loss (dissipation), unless the motion occurs in absolute vacuum, which is a theoretical abstraction. The general premise of studying viscous friction at micro- and nano-scales in the continuum assumption has been Stokes' drag equation[16,17], which states that the damping losses are proportional to the viscosity of the fluid when the other flow parameters such as velocity and length scale are held constant. However, this law of friction starts showing deviations at the nanoscale, prompting the need to understand the role of viscous interactions at nanoscale interfaces arising from the solid-fluid interactions[8,18–35].

To elaborate on the flow of fluid at the nanoscale in the Stokes flow regime, the physical length scale $a$ is of the orders of nanometers and the typical velocity $u$ is of the order of mm or µm per sec. In such systems, viscous forces dominate over inertial forces, as expressed by the dimensionless Reynolds number $\Re e = \dfrac{ua}{\nu}$, $\nu$ being the media's kinematic viscosity, with $\Re e \ll 1$, and where $\nu = \dfrac{\eta}{\rho}$, the ratio of $\eta$, the dynamic viscosity and $\rho$, the density of the fluid. For uniformity in all further discussions the term viscosity has been used to mean kinematic viscosity $\nu$. This very low Reynolds number regime is most relevant for vibrating nanomechanical oscillators or surfaces modified with nanoscale structures, where the effective flow is across nanostructure boundaries. Since the viscous terms dominate the inertial terms in this Stokes regime, the equation of motion shows a linear dependence on velocity, and the dissipation terms are proportional to viscosity. For an incompressible fluid motion without external forces, the Stokes equation is

$$\nu \Delta \mathbf{u} - \frac{1}{\rho}\nabla p = \mathbf{0},\tag{1}$$

with $\nabla \mathbf{u} = 0$ inside the fluid and $\mathbf{u} = 0$ on the solid-fluid interface, where $\Delta$ is the Laplace operator, $\rho$ is the fluid density assumed constant and $p$ is the pressure. In this approximation, the viscous drag on any finite object moving through the fluid will be proportional to the velocity of the object and the viscosity, with a coefficient depending on the object's shape. It should be noted that the shape coefficient may depend weakly on $\mathfrak{R}e$ due to Oseen's corrections as we shall utilize below for the case of a cylinder moving through the flow. Nevertheless, the main dependence of the friction force acting on an object in Stokes' approximation must be linear (or almost linear) in viscosity and velocity. The effect of viscous drag in the Stokesian regime can be studied by the measure of damping experienced by a mechanical resonator, where the motion generates oscillatory flows at the interface. The damping can be characterized by the Q-factor of the resonator, proportional to the inverse of the rate of decay of vibrations per period. Energy loss from viscous drag reduces the Q of the system, which depends on the interactions of the resonator surface with the medium.

For liquids, the variations in viscosity can be significant, by several orders of magnitude. For gases, however, that variation is much smaller. This is because the mean free path of molecules in gases (~67nm in normal conditions) is large compared to the inter-molecular distances (~ 3 – 4nm under same conditions) and the interaction forces between molecules decay rapidly with increasing distance. The values for gas viscosity thus tend to be small and change relatively little as compared to liquids. In addition, the dynamic range of variation for kinematic viscosity between different gases is relatively small, within an order of magnitude. For example, viscosity of $H_2$ is $8 \mu Pa \cdot \sec$ while that of $CO_2$ is $14.8 \mu Pa \cdot \sec$. Viscosities of gases also show little variation with temperature[15].

This small dynamic range of gas viscosity limits observable variations in interactions with a surface, since its effect on dynamical change is correspondingly small. We shall also note that oscillating shear flows over a surface lead to power-law dependences of drag on viscosity, which is also a slowly varying function. Thus, viscosity is not used as a parameter in characterizing vapors or gas media and in most sensing applications. Instead, most resonator sensors in air rely on resonator frequency shift induced by mass loading as a reliable sensor parameter. A resonator with a high resonance frequency, operating in vacuum can detect single molecules in special circumstances. In general, atto-gram sensitivity can be observed with micro and nanoscale mechanical resonators in high vacuum. The higher loss in energy due to fluid drag at the nanoscale necessitates operation in vacuum to increase the Q of resonance in order to obtain a high resolution in frequency variations. However, it has been reported that the improvement in Q at higher vacuum for nano-cantilevers is rather moderate[7,18,20]. This would suggest that it is difficult to use the change of gas viscosity as a measuring tool, whether by a macro- or a nano-scale device.

**Objectives and key results of the work.** We report that in contrast to existing results for single oscillators at the nanoscale, modifying an oscillating surface with a forest of nanoscale features, like vertical slender nanorods or nanobristles (Fig. S1), magnifies viscous interactions with the ambient media by up to three orders of magnitude as compared to the non-modified (bare) surface. This was observed with surface nanostructures having typical spacing between them in the order of 50-60 nm (Fig. 1a), which is around the mean free path $(mfp)$ of the gas phase molecules. As far as we are aware of, the studies of enhancement of dissipation from such surface nanostructuring has not been addressed in the literature. More interestingly, we observe an enhancement showing an *exponential* sensitivity to viscosity change, drastically deviating from the linear dependence as predicted by the premise of Stokes' theory and its modifications.

**Experimental design.** A schematic of the experimental realization (Fig. 1a) in the context of studying such surface nanostructuring effect in a quartz crystal (QC) oscillator is shown along with the key result (Fig. 1b). Use of macroscopic, commercially available QC oscillators offers many advantages such as easy readout, simplicity, high base Q-factor in air, large surface area for increased molecular interactions, and low cost. The nanoscale modification of QC surface combines the advantages of both nano and macro devices and serves as a bridge between the nano and macro world. The closely spaced nanostructures increase the mechanical interaction between gas molecules and surface, leading to an exponential dependence on viscosity, as we report.

In contrast, the change observed for an unmodified bare surface is consistent with the Stokes' equation, and is orders of magnitude smaller. For small variations in the kinematic viscosity, Stokes' theory predicts the linear change (see schematic on Fig. 1c). Our measurements of the response of the bare QC surface (Fig. 2b) attests to solutions of Stokes' equation for an oscillating surface (Stokes' second problem[36]), which predicts damping rates $D$ as slow varying nonlinear functions of media properties. Typically for shear flows, $D$ scales as $\nu\rho$ where vibrations generate thick viscous layers[24,30,34], or as $\rho\sqrt{\nu}$ when the viscous layers are thin[9,19,36]. In the Stokesian regime in air, the viscous layer is thin and hence the relative change in $D(\nu,\rho)$ can be approximated with a high accuracy as $\frac{\Delta D}{D} \cong \frac{1}{2}\frac{\Delta \nu}{\nu} + \frac{\Delta \rho}{\rho}$ for small (few per cent) relative changes in $\rho$ and $\nu$, relevant to media characterization. For more general functions of $D(\nu,\rho)$ involving higher order terms, the changes in $\frac{\Delta D}{D}$ would still remain linear in small relative changes of parameters. More complex theories based on Stokes' flows can be derived, and they all point to a *linear* dependence of dissipation on small relative changes in the ambient media. This contradicts our results for nanostructured resonators which we will analyze below in detail.

**Measured quantities and presentation of results.** All our results are presented in terms of the dissipated to stored energy per cycle at resonance as

$$D = \frac{1}{Q} = \frac{E_{diss}}{E_{stored}} = \frac{\int_0^T \vec{F}_{visc} \cdot \vec{\dot{x}} dt}{E_{kinetic} + E_{potential}} = \frac{\int_0^T \gamma \dot{x}^2 dt}{\frac{1}{2T}\int_0^T (\dot{x}^2 + \omega_0^2 x^2) dt} = \frac{\gamma \int_0^T \omega_0^2 x^2 dt}{\frac{1}{T}\int_0^T \omega_0^2 x^2 dt} = \frac{\gamma}{f}, \quad (2)$$

from simultaneous measurements of $f$ and $\gamma$ in an experiment (Fig. 1a) where $\vec{F}_{visc} = \gamma \dot{x}$ is the viscous drag force, $\gamma$ the damping factor and where $f = 2\pi/\omega_0$ is the resonance frequency in sec$^{-1}$, $T$ being the time period of oscillations. The motion in this case is assumed to be along the $x$ direction with $\dot{x}$ and $\ddot{x}$ denoting the 1$^{st}$ and 2$^{nd}$ order time derivatives respectively.

We note here, that the obtained $D$ from experimental impedance analysis and $\frac{1}{Q}$ obtained independently from the Lorentzian fit (see Fig. S3 in Supplementary materials) are identical with accuracy to an order of 10$^{-7}$ (Fig. S4 in Supplementary text). This serves as an additional verification of the accuracy of our measurements. Therefore, we can consider both $D$ or $\frac{1}{Q}$ as identical. It is easier to cast the results in terms of $D$ since it is easier to analyze the total dissipation in the system as the linear addition of internal dissipation and that originating from boundary media interactions.

**Experimental results of dissipation enhancement for nanostructured interface.** Our work reports an *exponential* sensitivity of dissipation for various gas media with a nanostructured resonator, which is drastically different from previous results. The enhancement effect was studied in detail by conducting systematic experiments for various media conditions (Fig. 2a), allowing a dynamic range of up to 30% variations in kinematic viscosities. The experimental kinematic viscosities for different media conditions were computed using literature reference and has been included in the supplementary materials (S5). Previous works report similar surface nanostructuring effects, attributing enhanced sensitivity to increased surface area for higher molecular adsorption and mass loading on an oscillator[38,39]. Our approach of dissipation analysis reveals a drastically different paradigm, namely, the exponential change in damping for even few percent relative changes in media kinematic viscosity $\nu$ (Fig. 2a). The data in Fig. 2 includes the change of viscosity corresponding to different gas media and also due to change of temperature[15] in the same gas media. A crucial result is the observed exponential dependence of dissipation on temperature of a fixed gas as well, as shown in Fig. 2a, with data collapsed on to the same normalized scale of kinematic viscosity change. For a detailed experimental setup and operation, refer to the section *Experimental Methods* below. All normalized x-axis scales represent relative changes of corresponding 'variable' as in $\left(\frac{\Delta \text{variable}}{\text{variable}_{\min}}\right)$. The observed enhancement effect of nanostructuring is shown to resolve viscosity changes substantially better compared to non-nanostructured surfaces at even normal atmospheric conditions (Fig. 2b, 2c), effectively broadening the detection range of the otherwise narrow range of viscosities of gases. In comparison, the resonance frequency variations of the resonator are orders of magnitude lower (Fig. 3). Error bars shown on all experimental data points represent 5% of absolute variation. It is also interesting that the change in damping is drastic for even modest changes in the resonance amplitude $a$ (Fig. 4), making the dissipation response strongly non-linear, even though the resonator itself is operating in the purely linear regime. For a standard linear resonator, the dissipation would have varied proportional to the resonance amplitude. In contrast, nanostructuring of the surface creates a drastic exponential decrease with the resonance amplitude.

We note here that the dissipation has contributions from resonator's internal friction, and friction with external media as $D = D_{\text{int}} + D_{ext}$. It is known that friction with external media ($D_{ext}$) depends on both shape factor and media-boundary interactions. For small amplitudes, $D_{ext}$ is the dominating factor for a nanostructured resonator surface as revealed in the experimental results (Fig. 4). However, for higher resonance amplitudes, the boundary-media dissipation component becomes small, and the internal dissipation $D_{\text{int}}$ can no longer be neglected, exhibiting a residual dissipation in our system. This residual dissipation has been included in the exponential fits of Fig. 4 of all presented data related to the amplitude dependence. In the enhancement case shown on Fig. 2, this residual dissipation $D_{\text{int}}$ is negligible compared to the boundary-media dissipation term.

**Limiting Sensitivity: Relevance to sensing applications.** Our work shows that nanostructuring a surface allows probing the most fundamental physical aspects of ambient gases, such as its continuity, using a macro-device, acting as a bridge between the nano and macro world. Such active non-continuum probing resolves the changes in viscosity better for magnified viscous effects at the interface, effectively broadening its dynamic range (Fig. 2b). We show experimentally that such enhanced resolution cannot be achieved with non-modified surfaces (Fig. 2c).

It is imperative to provide an estimate of the maximum possible sensitivity achievable by a measuring device, based on such nanostructuring, using dissipation as an observable parameter. Such an estimate is most relevant for cases of a non-adsorbing analyte in very low concentrations such as highly volatile organics. As the data shows (Fig. 3), using frequency as a measuring parameter for the sensor in normal conditions is challenging in such a case as it is orders of magnitude smaller and does not follow any established trend for small changes in media. It is also known that for dominant boundary-media viscous interactions the resonance frequency is not independent of the damping forces. So, deducing changes in the media through resonator's dynamics solely from the aspect of resonance frequency shift can be difficult where the adsorption effects are small. In our work, for high enough changes in relative humidity, the observed change in the resonance frequency is in the parts per million order conforming to previous works[38,39]. On the other hand, damping shows the relative change of several orders of magnitudes, making the damping in surface nanostructured resonators to be approximately $10^8 - 10^9$ times more sensitive to ambient changes compared to the frequency shift (*i.e.,* factor of $10^2 - 10^3$ in dissipation vs $10^{-6}$ in frequency).

To elaborate on this point, the changes in dissipation are expected to come from the probability of molecular interactions and fluctuations in a cell bounded by the nanostructure walls, and this method does not necessitate adsorption of the molecules on the resonator to effect a change in resonator's behavior. Let us consider the volume space, marked in yellow in Fig. 5, contained in-between adjacent nanorods (cell), and suppose $\alpha$ is the concentration of an added analyte expected to induce a change in the perceived media. Then, at each point in time, the probability that a cell would have an analyte molecule in it is $^1P = \alpha$, and not to have that molecule is $^0P = 1 - \alpha$. After a characteristic viscous timescale $\tau_m = L^2/\nu$, the molecules of the analyte will escape the cell, making the events that are separated by $\tau_m$ time apart statistically independent. After an elapsed time $T' = n\tau_m$ encompassing $n$ independent events of going in/out of a cell (equivalent to one event from $n$ independent cells), the probability of a cell not to experience the effect of presence of an analyte molecule will be $^0P_n = (1-\alpha)^n = e^{-\alpha T'/t_m}$ assuming $\alpha \ll 1$ and $n \gg 1$. For a typical time scale of one QC period at resonance, $T' \Leftrightarrow \tau_n = 1/f$, the number of cells that are actively experiencing the effect of analyte molecules becomes $N_s = N_r\left(1 - e^{-\alpha \tau_n/\tau_m}\right) = N_r(\alpha \tau_n/\tau_m)$ for $\alpha \to 0$, with $N_r$ being the total number of cells per unit area. The limit sensitivity of the system, as measured by concentration of molecules, is then given as $\frac{N_s}{N_r} = \alpha \frac{\tau_n}{\tau_m}$. With the current experimental apparatus, measurement accuracy of $D$ is guaranteed to be better than 0.02% which gives $\alpha \approx 2 \times 10^{-5}$ (~10 parts per million - ppm) for $\tau_n/\tau_m \approx 10$ typical to our system. However, the accuracy of the measurement of $D$, as shown in the Supplementary Section S3, fits the alternative measurement of the width of Lorentzian within $10^{-7}$, (Figures S3 and S4), giving the expected minimum concentration measurable with this setup to be $10^{-8}$ (~10 parts per billion - ppb), if one were to assume that the accuracy of the measurements illustrated in Figures S3 and S4 persisted for all experiments. In principle, the limit of the accuracy is given by mean average fluctuation of number of molecules in a cell (about $\pm 3\%$ for $10^3$ molecules in a cell between nearest neighbor rods), taken for $N_r = 10^8$ independent cells $/mm^2$, giving an estimate of $\alpha \approx 10^{-9}$ (ppb). Of course, such idealistic experiment to perceive variations in ppb level will require much more accurate measurements from the aspect of electrical impedance accuracy, though not unachievable; and may require controlled conditions such as low temperature *etc*. This limiting value of measurable concentrations is of great importance for design of devices that can achieve ppb accuracy even at normal conditions.

**Theoretical analysis of the results.** We shall now focus on a theoretical understanding of the system to explain the experimental results obtained. For a plausible explanation of the exponential dependence of the dissipation on parameters, we suggest the following theory based on dynamical transitions between two co-existing states at resonance, in-phase and out-of-phase motion of nanostructures with the motion of the resonator base. We base this theory on the assumption that most of the dissipation comes from the out-of-phase motion, with the transition between the states occurring due to the stochastic transitions between the multiple steady states and Stochastic Resonance (SR) effect[40–42]. Indeed, such a consideration of stochastic transition is relevant since the nanostructure dimensions and distances between them are smaller than the *mfp* of the gas phase molecules.

*State I: In-phase motion of nano-structures.* It is well known that the resonator motion would generate oscillating flows at the interface, *i.e.* nanostructure boundaries in our case, with $\Re e \sim 10^{-6}$ and the corresponding dynamic viscous length scale[43] $\delta = \sqrt{\dfrac{2\nu}{\omega_0}} \sim 10^{-6} m$, which is much greater than both the width $w \sim 50 - 60 nm$ between a nearest neighbor nanostructure, and also the *mfp* of the gas molecules. Thus, due to shear flow, the viscous interaction from each nanorod would extend laterally over many neighboring nanorods on the surface, resulting in a strong bias towards the in-phase steady motion of the nanorods with the base (Fig. 5a), similar to swimming microorganisms[44,45], clustered stereocilia during spontaneous oscillations of hair bundles in cochlear outer hair cells[21], or Kuramoto oscillators[46]. The resulting viscous friction between the rods from the oscillating-flow would then account for the energy loss in the form of damping in experiments. This loss energy scale can be estimated as the work of friction force per period of oscillation[43], given by $\Delta W = 2\pi \nu \rho C L u^2 / f$, where $C = 2/\log(7.4/\Re e)$ is the Oseen's drag coefficient correction for low $\Re e$ flow past a rigid cylinder[43] with $L$ and $d$ as defined in Fig. 1a and $u$ being the characteristic flow velocity. This energy dissipation expression essentially leads to Stokes-like drag formulations[10,18,20,27,29,37,47], predicting a *linear* dependence of dissipation $D$ on the small relative changes in viscosity $v$. This predicted result is expressed graphically in Fig. 2a by the gray dash-dotted line highlighting the 2 orders of magnitude deviation from experimental results. We also studied more complex considerations of Stokes flows with off-set boundaries to account for the nanostructured surface, leading to a similar, essentially linear estimate of resonator friction dependence on viscosity. However, our experimental results deviate significantly from this Stokesian regime. Thus, there must be another state of the system with drastically enhanced dissipation. This other state, at resonance, we can relate to with the out-of-phase motion of the nanostructures.

*State II: Out-of-phase motion.* Because of inherent randomness in the system, a spontaneous off-phase motion of some neighboring pairs of nanorods generating flow transversal to the resonator plane may occur (Fig. 5b). With first order approximation, let us consider the volume space $V = Lw^2$ bounded by counter-moving walls of the nanorods (cell) (Fig. 5). The in-phase moving walls of the cell do not change the volume contained between them and thus generate no additional flow velocity. The out-of-phase motion leads to an effective volume change between two such nearest neighbors as $dV/dt = U \cdot S_\perp$, thus generating velocity $U$ of the gas, with $S_\perp = w^2$ being the cross sectional area of the cell normal to the surface of the crystal (Fig. 5), $w$ being the distance between rods. This would account for localized energy of the fluid moving in the out-of-phase cell as

$$\Delta E = V \cdot \rho \cdot U^2 / 2 ,  \qquad (3)$$

$\rho$ being the gas density, assumed constant for incompressible flow. For the typical dimension of the nanorods and operational vibrational amplitude $a \sim 4nm$ (See Supplementary Section S7), the mean free path of the molecules is $l_* \cong 67nm >> a$. Since the resonance amplitude is comparable with the intermolecular gas distances $(3-4nm)$, there is a high rate of impact between the molecules and the nanostructures. In addition, the impacted molecules are much more likely to collide with another nearest neighbour nanorod than another molecule. We can approximate this fact by positing the effective amplitude of motion of molecules to be $a_{eff} \cong l_*$. With this consideration, we can approximate the time rate of volume change for off-phase motion as

$$dV/dt = S_\| \cdot f \cdot a_{eff} = L \cdot w \cdot f \cdot a_{eff}, \qquad (4)$$

$S_\| = Lw$ giving the cross-sectional area in the direction parallel to the crystal surface (Fig. 5) and $f$ being the frequency as defined before. This time rate of change of the volume may cause the gas to move in and out of the domain bounded by the walls with a typical velocity $U = \dfrac{dV/dt}{S_\perp} = L \cdot f \cdot a_{eff} / w$. It thus follows from equations (3) and (4), that the energy gain necessary to go from in-phase to out-of-phase motion with the base is $\Delta E = \rho L^3 a_{eff}^2 f^2$. As the distance between the rods is comparable to molecular $mfp$, the lubrication-type friction caused by impacts between the molecules can be considered small compared to that arising from the stochastic interactions with the walls. Thus, the main contribution to dissipation can be assumed to be originating from the friction encountered by the rods themselves[33,43].

*Transition between the states due to stochastic dynamics.* The existence of these two states, representing the two possible steady states of the system at resonance, can be considered as the first step towards a more complex theoretical analysis. In reality, there is a continuum of states corresponding to the intermediate configurations of flow. The states can be viewed in terms of effective variable $\phi$ representing the on and off-phase transition of an individual cell, with $\phi = 0$ being the in-phase state and $\phi = \pi$ being the out-of-phase state. Then, $\Delta E$ is the energy barrier the system needs to overcome in order to transition from the in-phase to the out-of phase state, as schematically illustrated by a double-well potential on Figure 5, the minima of the potential representing the two possible states. The transition between the states is driven by the molecular impacts which are much faster (~ picosecond) compared to both the time scale of oscillations of the crystal (~ microsecond) and the resonance of the nanorods (~ nanosecond). The dynamics of the transition between the states can then be modeled in terms of stochastic dynamics of a bi-stable system[40–42] with the noise intensity $\xi$ (i.e., the noise autocorrelation function) being proportional to $\Delta W$, which is precisely the viscous loss due to the molecules impacting the rods.

**Key result on the Theoretical Analysis and relevance to experimental data.** For detailed theoretical analysis to be undertaken later, one will proceed with finding a stochastic dynamics model of motion between the different states. Such a model will necessarily include the dynamical assumptions of the state $\phi$ as a function of generalized time, the potential $\Psi(\phi)$ describing the energy of a given state, a periodic forcing and a noise. Such theory will be crucial for understanding of the dynamics close to the resonance, however, for now, we would like to postpone such detailed discussions for subsequent work when the nature and functional form of different terms is more clear. Here, we take a more general approach and outline the results that will be common for such models, and which are of relevance to our experiments. The primary ingredient of the theoretical model is the potential function $\Psi(\phi)$ defining the energy of a given state, having the minima at $\phi = 0$ and $\phi = \pi$ with the potential barrier separating the two states being $\Delta E$ (Fig. 5) as we have illustrated Then, independent of the exact functional form of the chosen $\Psi(\phi)$, the noise-induced hopping between the states may be described in the form of Kramers rate[48,49]

$$r_K \propto \exp\left(-\frac{\Delta \Psi}{\xi}\right), \tag{4}$$

as long as the potential $\Psi(\phi)$ satisfies the above constraints of being a smooth function with two given minima. Surprisingly, this general fact is sufficient to provide an excellent explanation of all available experimental data. Indeed, with the majority of the states being in the in-phase state, Kramers' law, coupled with the assumption of a finite lifetime of the off-phase state, leads to describing the number of cells in the off-phase state following the Gibbs probability distribution as

$$p = p_0 \exp\left[-\frac{\Delta E}{\Delta W}\right] = p_0 \exp\left[-\frac{L^2 f}{4\pi C \nu}\left(\frac{a_{eff}}{a}\right)^2\right]. \tag{5}$$

This is precisely the consequence of the existence of two states as estimated above, with $\Delta W$ playing the role of the amplitude of the noise $\xi$ in (4). The exponent in square brackets above incorporates parameters of the integrated multi-scale resonator into a single dimensionless constant $K = \frac{1}{4\pi}\frac{L^2 f}{C\nu}\left(\frac{w}{a}\right)^2$ where $a_{eff} \approx w$, $a_{eff}$ being the effective amplitude of motion of molecules on impact with the rods as described before.

Now, assuming that most dissipation comes from the in-phase energy scale, with the off-phase energy scale resulting in the enhancement as a function of minute media viscosity $\nu$ variations, from (5) above we obtain

$$\Delta D = D_0 \exp\left(\frac{1}{4\pi}\frac{L^2 f}{C\nu}\left(\frac{w}{a}\right)^2 \cdot \frac{\Delta \nu}{\nu}\right) = D_0 \exp\left(K \cdot \frac{\Delta \nu}{\nu}\right), \tag{6}$$

with $D_0$ being the dissipation for the smallest kinematic viscosity in the experiments. While much needs to be done to derive a complete stochastic dynamics model from the first principles, we believe that this preliminary result coming from rather general assumptions leads to some encouraging insights. Indeed, the theoretical prediction (6) is in agreement with the exponential experimental trend presented on Fig. 2a and 2c, providing an excellent match to the data with no fitting parameters. The best fit of the graph in Fig. 2a provides $K = 9.15$ and theoretical prediction (6) yields $K \cong 9.1$ where we have taken $a = 4nm$ from experimental data analysis (supplementary materials S7). The high value of the dimensionless parameter K determines the amplification or scaling factor $e^K$ for unit changes in $\frac{\Delta \nu}{\nu}$, showing exponential sensitivity to small variations in viscosity in our experiments. In the absence of surface nanostructuring, the system ceases to have a second stable state where expressions (4) and (5) are no longer valid, and a continuum energy scale similar to $\Delta W$, relevant to the system, can explain the linearity in response for small relative variations in flow parameters[10,20,37]. This is confirmed by our experiments (Fig. 2b) where indeed, for a bare surface, the motion is in the continuum regime and expected change in dynamics from Stokes' drag is linear, $\Re e$ being a slow varying function on media kinematic viscosity $\nu$.

The theoretical prediction (6) for small changes in motion is further substantiated by the exponential dependence of dissipation on resonance amplitude (Fig. 4). If we assume the simplest linear dependence of $a_{eff}$ on $\Delta a$, from (6) we arrive to the exponential dependence of $D$ on $\frac{\Delta a}{a}$ (supplementary materials S8), shown with the dashed line as an experimental best fit in Fig. 4. Here $a$ in the denominator is the resonance amplitude obtained for the lowest input drive energy. More sophisticated models of dependence of $a_{eff}$ on $\Delta a$ with higher order terms can be derived, essentially leading to the same results. The most interesting fact is the apparent higher exponential rate of change of dissipation as a function of $\frac{\Delta a}{a}$ for the surface with nanostructures having diameter and width lesser than the *mfp* of the gas phase molecules (Fig. 4 Crystal 1). We shall note that the system with larger nanorods separated by a larger distance experience slower decay of $D$ with the amplitude (Fig. 4 Crystal 2). While the experimental limitations do not allow us to test the continuously changing dimensions of the nanorods, we believe that these results illustrate that, as the dimensions slowly increase with respect to the *mfp*, the effect of dissipation enhancement correspondingly diminishes, eventually yielding a linear dependence on viscosity as expected by Stokes' theory (Fig. 4a). Thus, even preliminary analysis for our system based on quite general assumptions are of substantial interest, and further elaboration of the theory will be considered in details in our upcoming work.

**Discussion on time and length scales.** We shall note that, the probabilistic consideration of interactions at the solid-gas interface, because of the nanostructures, inherently incorporates two non-dimensional scaling parameters: a time scale ratio $\frac{L^2 f}{\nu} = \frac{\tau_m}{\tau_n} = \Omega$ and the other a spatial scale ratio $\left(\frac{a_{eff}}{a}\right)^2 = \Pi$. The orders of magnitude of these two scale ratios gives a measure of separation of the two energy state basins considered in our model. Here, $\tau_m$ is the molecular motion time scale related to the kinematic viscosity $\nu$ and $\tau_n$ is the time scale of motion of the surface or the nanorods. The ratio $\tau_n/\tau_m = 1/\Omega > 1$ signifies the extent of collisions with gas molecules experienced by a nanorod per period of its motion or the rate of molecular fluctuations within a cell bounded by nearest neighbor nanorods. We have, typically, $\Pi \gg 1$ for our system, that changes depending on the grafting density on the surface. Their product $K = \Omega\Pi$ (~10 for typical order of magnitudes in our system) affords a large change of $p$ in (5) with respect to small changes in media defining the sensitivity. Theoretical estimate based on the physics of statistical fluctuations of the number of molecules in a cell bounded by the nanorods gives limiting sensitivity of the order of parts per billion (ppb) concentrations even at normal temperature and pressure conditions, on the introduction of an analyte, which can in principle be achieved by this method. Our experimental results demonstrate measurement sensitivity of the order of ppm as presented (Fig. 2a, 2c), when a single analyte is introduced in the system with a carrier gas.

**Conclusions.** Damping in nanomechanical resonators has traditionally been regarded as an impediment to sensitivity. We show that for oscillating surfaces modified with nanoscale structures, dissipation offers a wealth of information on the nature of mechanical interactions of molecules with surfaces. We also present a theoretical model based on two bi-stable states and two energy scales, showing encouraging agreements with experiments. We envision that a future development of theoretical studies incorporating ideas of Stochastic Resonance will be of particular interest. We believe that in the future, the analysis of dynamics of multitude of coupled nanostructures in complex gas mixtures may play an important role in non-adsorption based, physical detection of chemicals.

## Materials and Methods

**Surface Nanostructuring.** In our experiment, we use a standard AT cut QC from International Crystal Manufacturing (ICM) Co. Inc. (USA) as a test platform. First step involves sputter coating a $10nm$ thick ZnO seed layer only on one electrode surface using a mask, the coated area contributing to the effective response change in our case. We used hydrothermal[38] process for growing the ZnO nanorods on the sputter coated surface. The involved chemicals are: Zinc nitrate hexahydrate ($Zn(NO_3)_2 \cdot 6H_2O$, 98%) and Ammonium hydroxide (28 wt% $NH_3$ in water), purchased from Sigma–Aldrich. The crystals are put in $Zn(NO_3)_2 \cdot 6H_2O$ solution with its pH being modified to 10.6 by adding 2.3 ml of the ammonia solution. The solution is put in an oven at 90°C for 3hrs and the pH of the solution is constantly maintained by a sealed environment reaction vessel. The 3hr growth gave us our desired dimensions. After hydrothermal growth, the crystals were rinsed with de-ionized water and ethanol, and then dried in a vacuum oven. Estimations verified through Field Emission Scanning Electron microscopy (Fig. S1) show a mean nanorod length $L$ of $611nm \pm 10\%$ with mean diameters $d$ of the order of $43nm \pm 10\%$. The dimensions and growth kinetics strongly depends on the seed layer thickness and average roughness. Nanorod density measured at different locations is $18 \pm 2$ nanorods per $\mu m^2$ giving a typical area coverage of about 12% per $1\mu m^2$ of electrode surface area. Such a coverage density is essential to effect limit sensitivity ~ parts per billion as has been discussed in supplementary information.

**Flow system for analyte introduction.** A schematic of experimental procedure as shown (Fig. 6) allows varying of ambient media properties in the flow cell by the introduction of various gas molecules using carrier gas (dry-air) bubbling, with flow maintained at 50 sccm - standard cubic centimeters per minute for all cases. The variance in the media properties is achieved through 0.1 wt% analyte-water mixtures or by controlled temperature variations in the same dry air environment. The relative humidity for all experiments with different analytes and temperature is maintained at 5%. For same percentage by weight analyte solutions in water and the fixed flow rate, the relative change in kinematic viscosity is unique to an analyte depending upon the relative vapor pressures. For calculations on mixture viscosities in experiments refer section in supplementary information.

**Impedance measurement.** The nanostructure modified QC is driven into shear mode vibration and the dynamic dissipation factor as in our analysis (Eq. 1) is measured as an electrical impedance parameter $\gamma = \frac{\wp_s}{X_s} = \frac{\wp_s}{1/\omega c_s} = \omega \wp_s c_s$, where $\omega$ is the drive frequency, $\wp_s$ is the equivalent series resistance and $c_s$ is the equivalent series capacitance. At resonance (Eq. 1) $D = \frac{\gamma_{max}}{f_{QC}} = 2\pi \wp_s c_s$ gives the measure of dissipated energy. The measurements are done employing Agilent 4294A Impedance Analyzer with nominal impedance accuracy of $\pm 0.08\%$ at 100Hz. Drive signal - $V_{rms} = 5mV$ from analyzer's internal oscillator is swept over a moderately low bandwidth (5 kHz) with the sweep time of 30 sec for viscosity dependence experiments. Variations of impedance parameters monitored as a function of the drive frequencies, reflect the equivalent impedance based mechanical dissipation in the system.

# REFERENCES

1. *Surface and Interface Science: Vol 5 & 6*. (Wiley-VCH Verlag & Co., 2015).

2. Adamson, A. W. & Gast, A. P. *Physical Chemistry of Surfaces*. (John Wiley and Sons, 1997).

3. Min, Y., Akbulut, M., Kristiansen, K., Golan, Y. & Israelachvili, J. The role of interparticle and external forces in nanoparticle assembly. *Nat. Mater.* **7,** 527–538 (2008).

4. Pelton, M. *et al.* Damping of Acoustic Vibrations in Gold Nanoparticles. *Nat. Nanotechnol.* **4,** 492–495 (2009).

5. Tam, C. K. W. The drag on a cloud of spherical particles in low Reynolds number flow. *J. Fluid Mech.* **38,** 537 (1969).

6. Gieseler, J., Novotny, L. & Quidant, R. Thermal nonlinearities in a nanomechanical oscillator. *Nat. Phys.* **9,** 806–810 (2013).

7. Li, M., Tang, H. X. & Roukes, M. L. Ultra-sensitive NEMS-based cantilevers for sensing, scanned probe and very high-frequency applications. *Nat. Nanotechnol.* **2,** 114–20 (2007).

8. Oden, P. I., Chen, G. Y., Steele, R. A., Warmack, R. J. & Thundat, T. Viscous drag measurements utilizing microfabricated cantilevers. *Appl. Phys. Lett.* **68,** 3814–3816 (1996).

9. Rodahl, M. *et al.* Simultaneous frequency and dissipation factor QCM measurements of biomolecular adsorption and cell adhesion. *Faraday Discuss.* **107,** 229–246 (1997).

10. Rodahl, M., Höök, F., Krozer, A., Brzezinski, P. & Kasemo, B. Quartz crystal microbalance setup for frequency and Q-factor measurements in gaseous and liquid environments. *Rev. Sci. Instrum.* **66,** 3924 (1995).

11. Spletzer, M., Raman, A., Wu, A. Q., Xu, X. & Reifenberger, R. Ultrasensitive mass sensing using mode localization in coupled microcantilevers. *Appl. Phys. Lett.* **88,** 254102 (2006).

12. Widom, A. Krim, J. Q factors of quartz oscillator modes as a probe of submonolayer-film dynamics. *Phys. Rev. B* **34,** 1403 (1986).

13. Xu, Y., Lin, J.-T., Alphenaar, B. W. & Keynton, R. S. Viscous damping of microresonators for gas composition analysis. *Appl. Phys. Lett.* **88,** 143513 (2006).

14. Cleland, A. N. & Roukes, M. L. Noise processes in nanomechanical resonators. *J. Appl. Phys.* **92,** 2758–2769 (2002).

15. Sutherland, W. LII. The viscosity of gases and molecular force. *Philos. Mag. Ser. 5* **36,** 507–531 (1893).

16. Stokes, G. G. On the effect of the Internal friction of fluids on the motion of pendulums - Section III. *Trans. Cambridge Philos. Soc.* **IX,** 8 (1850).

17. Rayleigh, Lord. LXXXII. *On the motion of solid bodies through viscous liquid*. *Philos. Mag. Ser. 6* **21,** 697–711 (1911).

18. Bhiladvala, R. B. & Wang, Z. J. Effect of fluids on the Q factor and resonance frequency of oscillating micrometer and nanometer scale beams. *Phys. Rev. E* **69,** 036307 (2004).

19. Chen, Y., Zhang, C., Shi, M. & Peterson, G. P. Slip boundary for fluid flow at rough solid surfaces. *Appl. Phys. Lett.* **100,** 1–5 (2012).

20. Karabacak, D. M., Yakhot, V. & Ekinci, K. L. High-Frequency Nanofluidics: An Experimental Study Using Nanomechanical Resonators. *Phys. Rev. Lett.* **98,** 254505 (2007).

21. Kozlov, A. S., Baumgart, J., Risler, T., Versteegh, C. P. C. & Hudspeth, A. J. Forces between clustered stereocilia minimize friction in the ear on a subnanometre scale. *Nature* **474,** 376–9 (2011).

22. Martin, M. J. & Houston, B. H. Gas damping of carbon nanotube oscillators. *Appl. Phys. Lett.* **91,** 103116 (2007).

23. Ortiz-Young, D., Chiu, H.-C., Kim, S., Voïtchovsky, K. & Riedo, E. The interplay between apparent viscosity and wettability in nanoconfined water. *Nat. Commun.* **4,** 2482 (2013).

24. Sader, J. E. Frequency response of cantilever beams immersed in viscous fluids with applications to the atomic force microscope. *J. Appl. Phys.* **84,** 64 (1998).

25. Saviot, L., Netting, C. H. & Murray, D. B. Damping by bulk and shear viscosity of confined acoustic phonons for nanostructures in aqueous solution. *J. Phys. Chem. B* **111,** 7457–7461 (2007).

26. Jeffrey, D. J. & Onishi, Y. Calculation of the resistance and mobility functions for two unequal rigid spheres in low-Reynolds-number flow. *J. Fluid Mech.* **139,** 261 (1984).

27. Tamada, K. & Fujikawa, H. The steady two-dimensional flow of viscous fluid at low reynolds numbers passing through an infinite row of equal parallel circular cylinders. *Q. J. Mech. Appl. Math.* **10,** 425–432 (1957).

28. Thompson, P. A. & Troian, S. M. A general boundary condition for liquid flow at solid surfaces. *Nature* **389,** 360–362 (1997).

29. Tuck, E. O. Calculation of unsteady flows due to small motions of cylinders in a viscous fluid. *J. Eng. Math.* **3,** 29–44 (1969).

30. Van Eysden, C. A. & Sader, J. E. Frequency response of cantilever beams immersed in viscous fluids with applications to the atomic force microscope: Arbitrary mode order. *J. Appl. Phys.* **101,** (2007).

31. Voisin, C., Christofilos, D., Del Fatti, N. & Vallée, F. Environment effect on the acoustic vibration of metal nanoparticles. *Phys. B Condens. Matter* **316-317,** 89–94 (2002).

32. Verbridge, S. S., Ilic, R., Craighead, H. G. & Parpia, J. M. Size and frequency dependent gas damping of nanomechanical resonators. *Appl. Phys. Lett.* **93,** 20–23 (2008).

33. Yamamoto, K. & Sera, K. Flow of a rarefied gas past a circular cylinder. *Phys. Fluids* **28,** 1286 (1985).

34. Yum, K., Wang, Z., Suryavanshi, A. P. & Yu, M. F. Experimental measurement and model analysis of damping effect in nanoscale mechanical beam resonators in air. *J. Appl. Phys.* **96,** 3933–3938 (2004).

35. Yakhot, V. & Colosqui, C. Stokes' Second Problem in High Frequency Limit: Application to Nanomechanical resonators. *J. Fluid Mech.* **586,** 249–258 (2007).

36. Schlichting, H. & Gersten, K. *Boundary layer Theory*. (Springer-Verlag, 1999).

37. Ekinci, K. L., Karabacak, D. M. & Yakhot, V. Universality in oscillating flows. *Phys. Rev. Lett.* **101,** 1–4 (2008).

38. Joo, J., Lee, D., Yoo, M. & Jeon, S. ZnO nanorod-coated quartz crystals as self-cleaning thiol sensors for natural gas fuel cells. *Sensors Actuators B Chem.* **138,** 485–490 (2009).

39. Anderson, H., Jönsson, M., Vestling, L., Lindberg, U. & Aastrup, T. Quartz crystal microbalance sensor designI. Experimental study of sensor response and performance. *Sensors Actuators B Chem.* **123,** 27–34 (2007).

40. Gammaitoni, L., Hänggi, P., Jung, P. & Marchesoni, F. Stochastic resonance. *Rev. Mod. Phys.* **70,** 223–287 (1998).

41. Badzey, R. L. & Mohanty, P. Coherent signal amplification in a nanomechanical oscillator via stochastic resonance. *AIP Conf. Proc.* **850,** 1675–1676 (2006).

42. Benzi, R., Sutera, A. & Vulpiani, A. The mechanism of stochastic resonance. *J. Phys. A Math. Gen* **14,** 453–457 (1981).

43. Batchelor, G. *Introduction to Fluid Dynamics*. (Cambridge University Press, 1983).

44. Elfring, G. & Lauga, E. Hydrodynamic Phase Locking of Swimming Microorganisms. *Phys. Rev. Lett.* **103,** 088101 (2009).

45. Golestanian, R., Yeomans, J. M. & Uchida, N. Hydrodynamic synchronization at low Reynolds number. *Soft Matter* **7,** 3074 (2011).

46. Kuramoto, Y. Cooperative Dynamics of Oscillator Community. *Prog. Theor. Phys. Suppl.* **79,** 223–240 (1984).

47. Martin, M. J. & Houston, B. H. Gas damping of carbon nanotube oscillators. *Appl. Phys. Lett.* **91,** (2007).

48. Kramers, H. A. Brownian motion in a field of force and the diffusion model of chemical reactions. *Physica* **7,** 284–304 (1940).

49. Hänggi, P., Talkner, P. & Borkovec, M. Reaction rate theory: Fifty years after Kramers. *Reviews of Modern Physics* **62,** (1990).

## Acknowledgements

This research has been funded by Canada Excellence Research Chair (CERC) for Oil-Sands Molecular Engineering at University of Alberta. V.P was also partially supported by the National Sciences and Engineering Research Council (NSERC) Discovery grant and University of Alberta Centennial Fund. We thank Dr. Charles Van Neste for discussions and technical assistance, Dr. Dongkyu Lee for discussions regarding fabrication of nanorods and Dr. Ravi Gaikwad for the SEM images.

## Author Contributions

A.P synthesized, characterized, designed and performed all experiments. J.E.H wrote machine interface codes in LabVIEW and was involved in study design and data interpretation, K.P was involved in study design and data interpretation, A.P, V.P, and T.T were involved in study design, data interpretation and theoretical analysis, and wrote the manuscript. All authors discussed results and commented on the manuscript.

## Additional information

**Supplementary information** accompanies this paper.

Correspondence and requests for materials should be addressed to A.P., V.P. or T.T.

**Competing financial interests:** The authors declare no competing financial interests.

**FIGURES:**

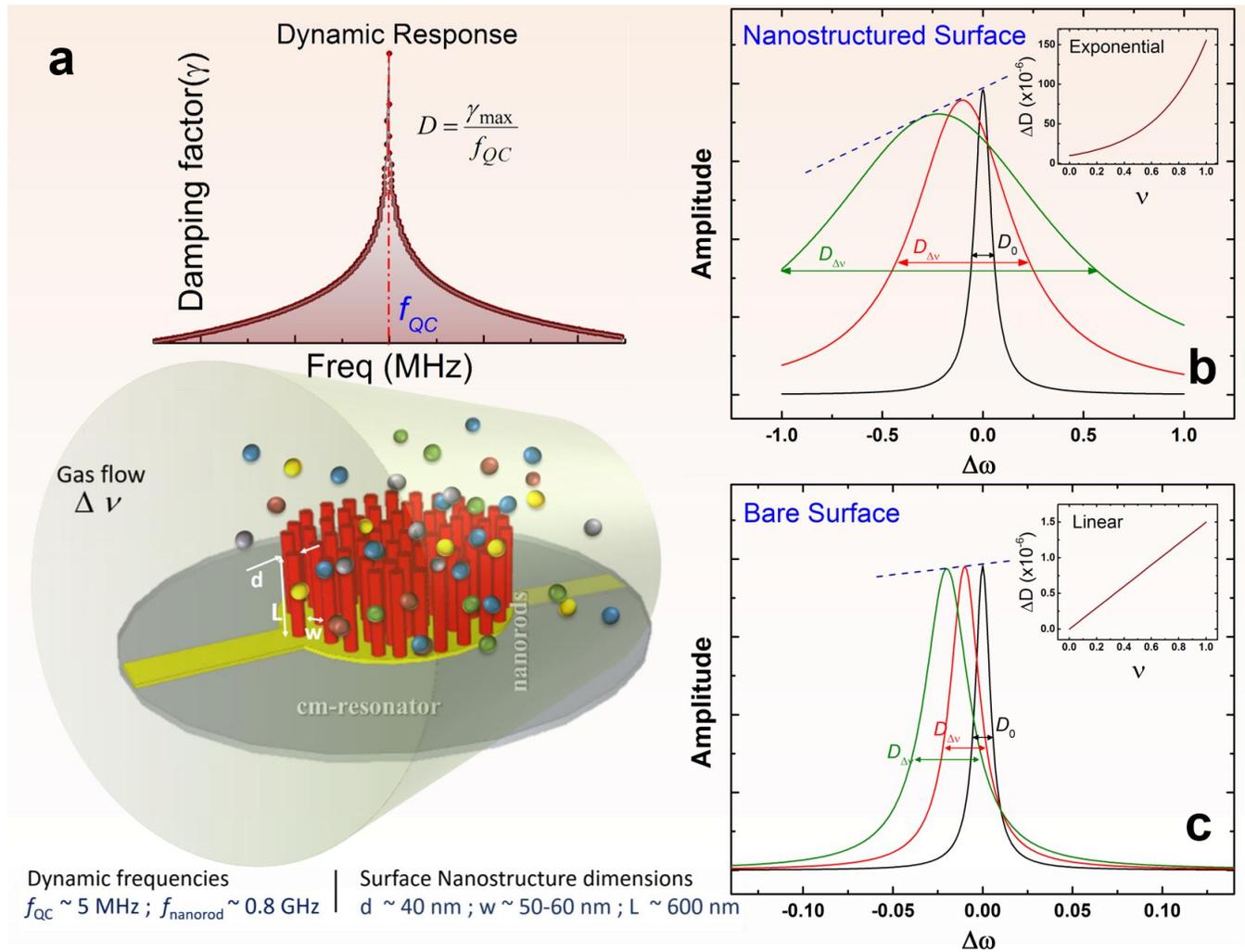

**Figure 1| Experimental Schematic with key result. (a)** A typical measured graph of dissipation (top) and an artist's impression of a nanostructured resonator surface interacting with gas molecules (bottom). **(b)** Amplitude response graphs for a nanostructured surface for three measurements of increasing viscosity of ambient gas (black, red, green) showing the exponential variation of the dissipation $D$ on viscosity (inset). **(c)** Amplitude response graphs for a bare surface for similar three measurements of increasing viscosity of ambient gas (black, red, green) showing the linear variation of $D$ on viscosity (inset). The blue dotted lines joining the amplitude peaks in **(b)** and **(c)** represent the same relative change in amplitudes in both cases.

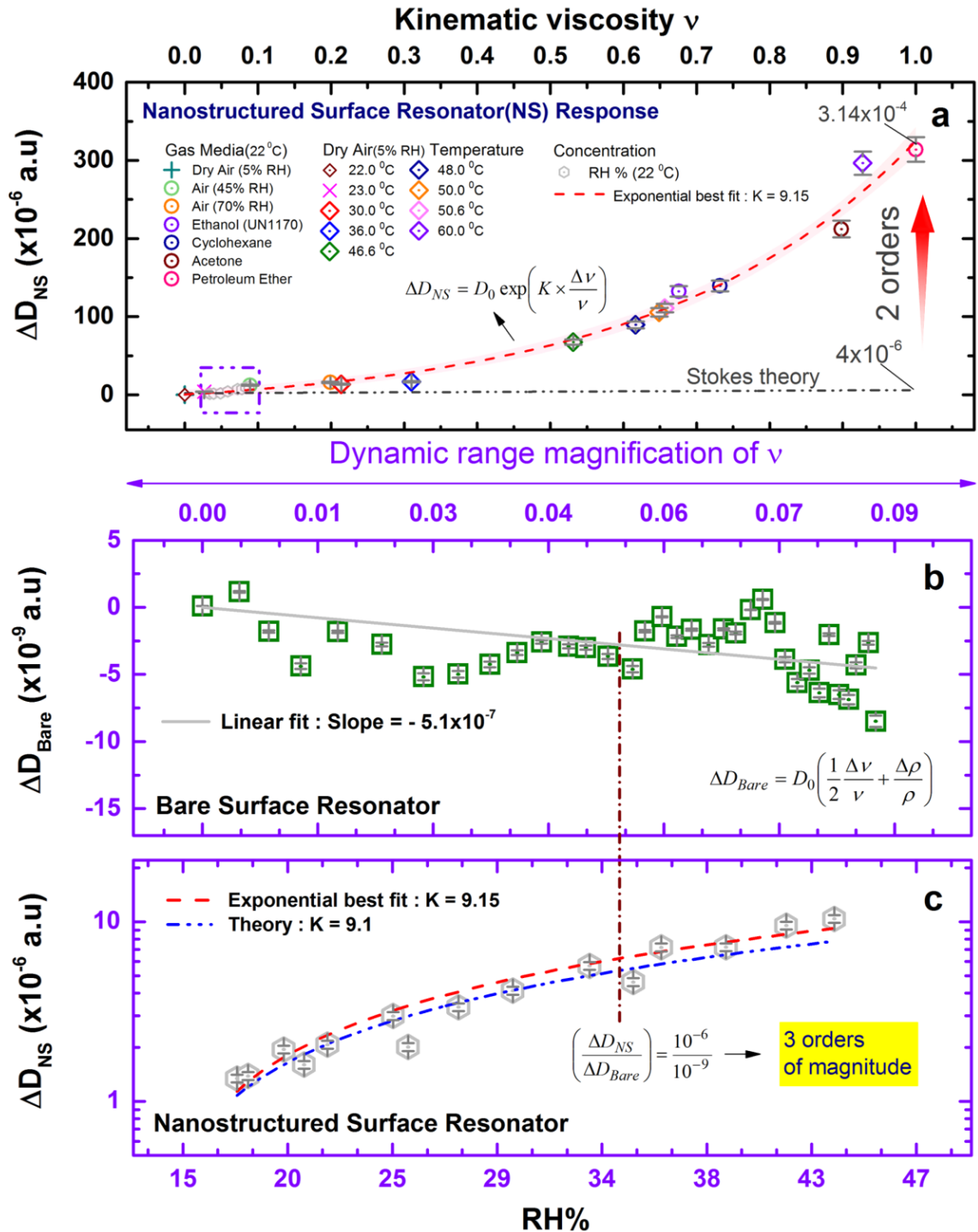

**Figure 2| Damping with varying media conditions. (a)** Results of experiments for a wide range of viscosities for different gases at a fixed temperature, and also temperature dependence for a single gas. 2 orders of magnitude deviation from Stokes' theory shown with a Red arrow. Panels **(b)** and **(c)** show the effective magnification of dynamic range of $\nu$ as a measurement parameter, where 3 orders of dissipation enhancement as compared to a bare surface are highlighted.

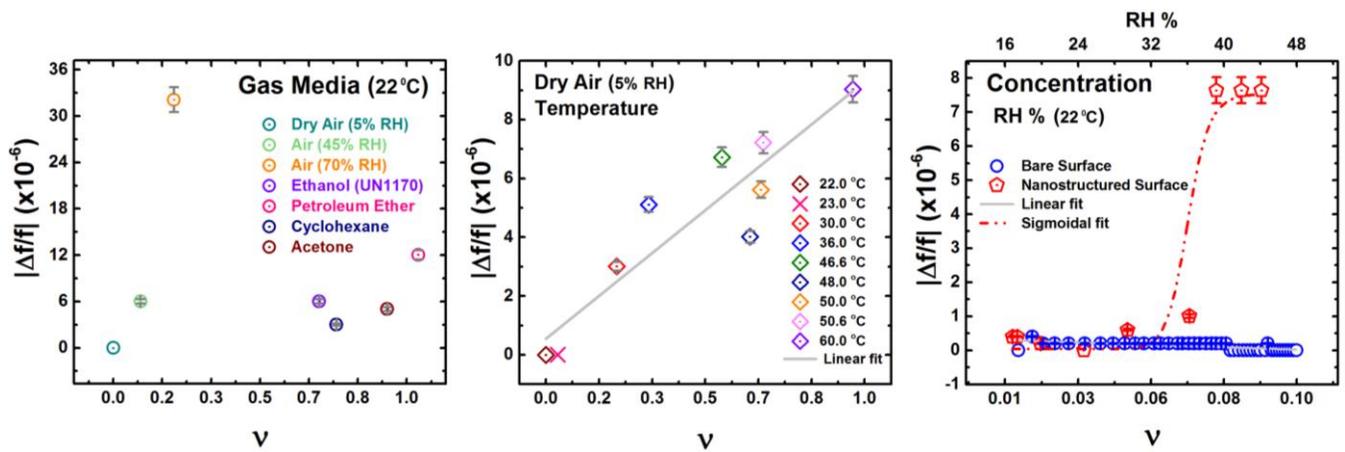

**Figure 3| Observed absolute relative frequency changes** $|\Delta f/f|$. Resonance frequency $f$ measured in dry air at normal temperature and pressure conditions for the lowest drive input. All variations in frequency correspond to the expected values with orders of magnitude of $10^{-6}$.

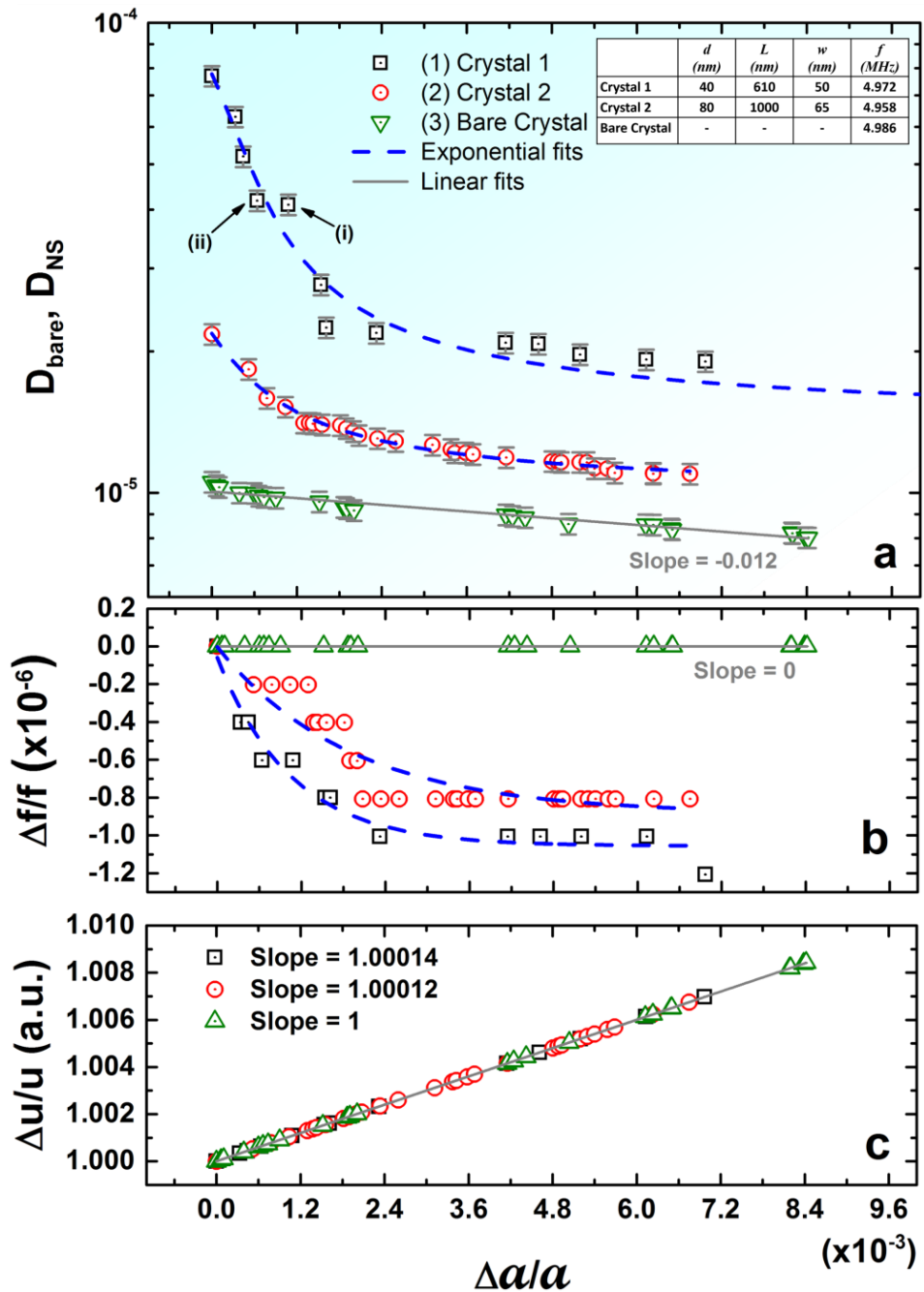

**Figure 4| Damping for a single vapor concentration vs normalized changes resonance amplitude.** (a) Shows high sensitivity of the dissipation for a nanostructured resonator vs an equivalent resonator with a non-structured surface. Dashed blue line shows exponential fit corroborating to our theory. As a representative for detailed analysis check Fig. S3 for data points (i) and (ii). (b) The corresponding frequency variations, commonly used as a measuring parameter, shows substantial enhancement for nanostructured case, and remains in the order of ppm. (c) The variations of velocity are linearly proportional to the change in amplitude, with the slope being very close to 1, as expected.

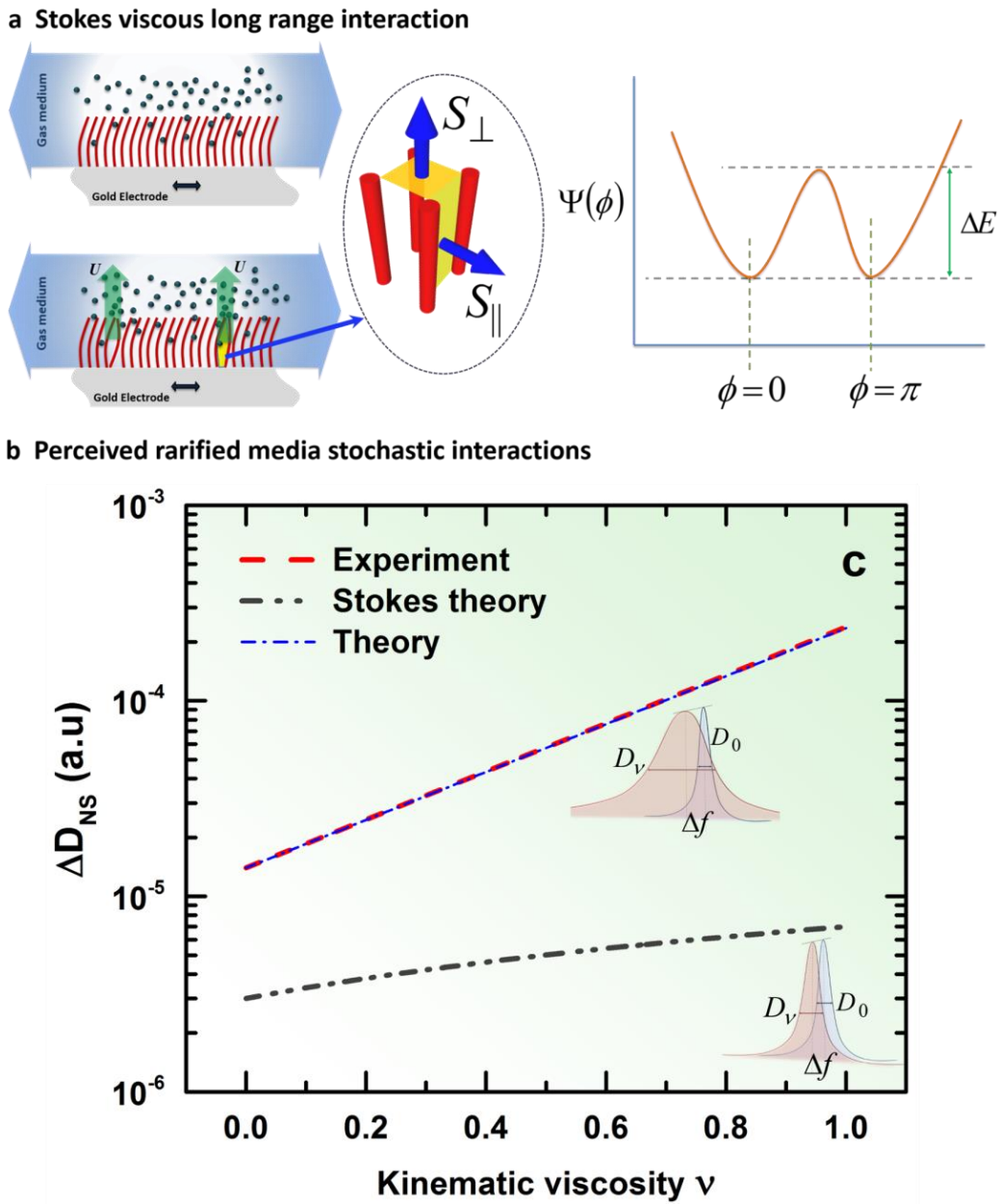

**Figure 5| Illustration of the mechanism for spontaneous out-of-phase rod dynamics enhancing dissipation.** (**a**) Illustrating in-phase motion of the rods, only the friction of the motion parallel to the resonator contributes to dissipation. (**b**) Two incidents of out-of-phase motion of the rods create a strong motion transversal to the resonator, enhancing the dissipation. The potential function $\Psi(\phi)$ defining the energy of a given state, having the minima at $\phi = 0$ and $\phi = \pi$ with the potential barrier separating the two states (the in-phase and out-of-phase) being $\Delta E$. (**c**) Compiled graphical comparison of experiment and theoretical analysis.

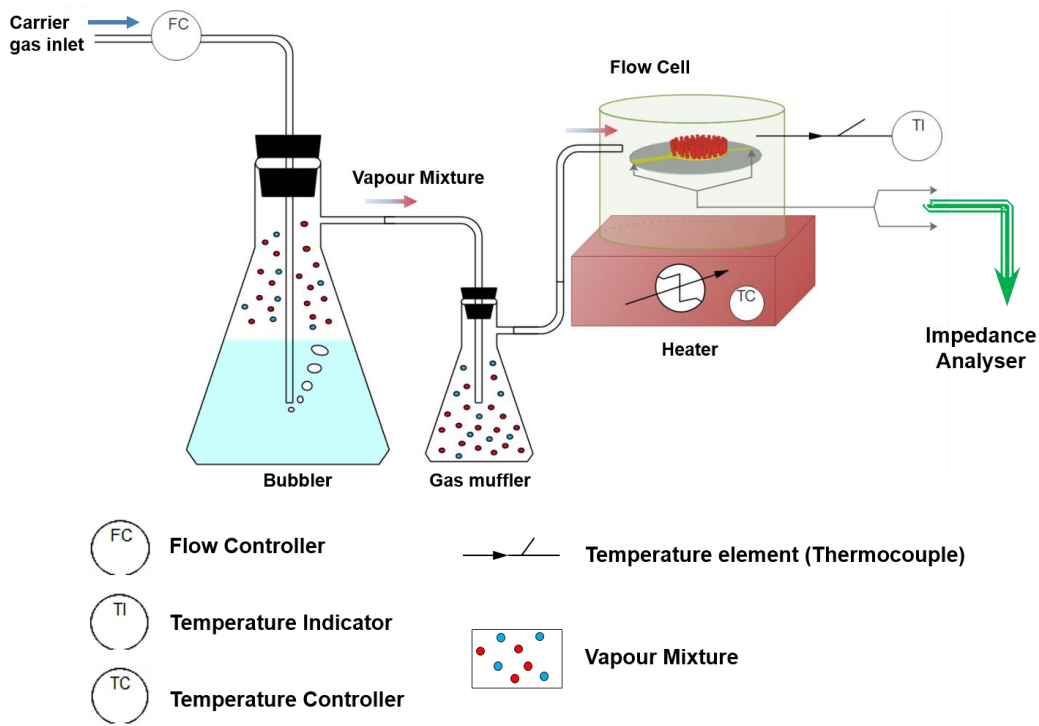

**Figure 6| Experimental Schematic** (Materials and Methods)